\documentclass[aps,prb,a4paper,superscriptaddress,twocolumn,floatfix,showpacs]{revtex4}
\usepackage{graphicx}
\usepackage{latexsym}
\usepackage[]{amsmath}
\usepackage{revsymb}

\begin{document}
\newcommand{\chem}[1]{\ensuremath{\mathrm{#1}}}

\title{Weak magnetic transitions in pyrochlore Bi$_2$Ir$_2$O$_7$}

\author{P.\ J.\ Baker}
\affiliation{ISIS Facility, STFC Rutherford Appleton Laboratory, Didcot OX11 0QX, United Kingdom}

\author{J.\ S.\ M\"{o}ller}
\affiliation{Oxford University Department of Physics, Clarendon Laboratory,
Parks Road, Oxford OX1 3PU, United Kingdom}

\author{F.\ L.\ Pratt}
\affiliation{ISIS Facility, STFC Rutherford Appleton Laboratory, Didcot OX11 0QX, United Kingdom}

\author{W.\ Hayes}
\affiliation{Oxford University Department of Physics, Clarendon Laboratory,
Parks Road, Oxford OX1 3PU, United Kingdom}

\author{S.\ J.\ Blundell}
\affiliation{Oxford University Department of Physics, Clarendon Laboratory,
Parks Road, Oxford OX1 3PU, United Kingdom}

\author{T.\ Lancaster}
\affiliation{University of Durham, Centre for Materials Physics, South Road, Durham, DH1 3LE, United Kingdom}

\author{T.\ F.\ Qi}
\affiliation{Department of Physics and Astronomy and Center for Advanced Materials, University of Kentucky, Lexington, KY 40506, USA}

\author{G.\ Cao}
\affiliation{Department of Physics and Astronomy and Center for Advanced Materials, University of Kentucky, Lexington, KY 40506, USA}

\date{\today}

\begin{abstract}

Muon spin relaxation measurements on Bi$_2$Ir$_2$O$_7$ show that it undergoes a bulk magnetic transition at $1.84(3)$~K. This is accompanied by increases in the muon spin relaxation rate and the amplitude of the non-relaxing part of the signal. The magnetic field experienced by muons is estimated to be $0.7$~mT at low-temperature, around two orders of magnitude smaller than that in other pyrochlore iridates. These results suggest that the low-temperature state represents static magnetism of exceptionally small magnetic moments, $\sim 0.01~\mu_{\rm B}$/Ir. The relaxation rate increases further below $0.23(4)$~K, coincident with an upturn in the specific heat, suggesting the existence of a second low-temperature magnetic transition.

\end{abstract}

\maketitle

Pyrochlore iridates are of interest in the field of frustrated magnetism because they offer a rare example where $5d$ effective spin $J=1/2$ magnetic moments decorate the pyrochlore lattice of corner-sharing tetrahedra and are subject to competing spin-orbit interaction, electron-electron repulsion, and crystal field energy scales.~\cite{rmpggg} Those pyrochlore iridates containing rare earths are described theoretically by focussing on the Ir subsystem,~\cite{pesin10,wan11,witczakkrempa12a,witczakkrempa13} the interactions between the $4f$ magnetic moments,~\cite{rmpggg,onoda10,savary12} and the $d-f$ exchange.~\cite{chen12xxx}
Previous work has led to predictions of interesting phases including a variety of topological insulators,~\cite{guo09,li10,pesin10,yang10} a topological semimetal with Fermi arcs,~\cite{wan11} and exotic quantum spin liquids,~\cite{savary12} neighboring more conventional metallic and magnetic phases.

The series of rare-earth pyrochlore iridates $A_2$Ir$_2$O$_7$ ($A=$ Y and Pr--Lu) offers the possibility of tuning the energy scales in order to access these theoretically predicted phases. Early experimental work found that with increasing ionic radius of $A$ the series goes from magnetic insulators~\cite{taira01} to unconventional metallic systems without magnetic ordering.~\cite{yanagishima01,maclaughlin10} 
In those systems where the iridium magnetic moments do order this is often,~\cite{matsuhira07,zhao11} though perhaps not exclusively,~\cite{disseler12nd} concomitant with a metal-insulator transition.

To separate the interesting properties due to the rare-earth and iridium ions, it is desirable to find compounds without a rare-earth magnetic moment or $f$-electrons, and a different ion size to yttrium. Bi$_2$Ir$_2$O$_7$ offers this opportunity. Having originally been noted for its metallic conductivity and small Seebeck coefficient~\cite{bouchard71} there was a long hiatus in magnetic measurements on Bi$_2$Ir$_2$O$_7$. 
This sits between the Eu and Nd pyrochlore iridates in terms of its lattice constant but the Bi$^{3+}$ ion has a larger ionic radius than any of the other reported pyrochlore iridates. It is therefore interesting to see whether any correlation can be found between the magnetic properties and either the lattice constant or the $A^{3+}$  ionic radius.
The recent study of Qi {\em et al.}~\cite{qijpcm} showed that the system remains metallic down to $2$~K and that there were no anomalies indicating any magnetic ordering in the magnetic susceptibility down to $1.8$~K, or specific heat down to $50$~mK. The specific heat exhibits an upturn below $\sim 250$~mK but since this continues down to $50$~mK it was not possible to determine whether this is associated with a magnetic transition or another phenomenon. The magnetic field dependence of the specific heat is also exceptional, with a strong field dependence of both the linear ($\gamma$) and cubic ($\beta$) contributions to the low temperature values.~\cite{qijpcm} While it is not uncommon for $\gamma$ to depend on magnetic field, $\beta$ is normally associated with the lattice contribution to the specific heat and this would not suggest any dependence on magnetic field. The resistivity shows a complicated temperature dependence and the Hall coefficient in a field of $7$~T changes from $\sim 2 \times 10^{-8}~\Omega$cm$^{-1}$ above $100$~K to $-1.6 \times 10^{-7}~\Omega$cm$^{-1}$ below $30$~K.~\cite{qijpcm} The results give an exceptionally large Wilson ratio $R_{\rm W} = \pi^{2}k^{2}_{\rm B}\chi/3\mu^{2}_{\rm B}\gamma = 53.5$, suggesting proximity to a magnetic instability or quantum critical point.~\cite{qijpcm}

Here we report a muon spin relaxation ($\mu$SR) study of Bi$_2$Ir$_2$O$_7$ extending down to $50$~mK. In zero applied magnetic field we observe two transitions in the muon spin relaxation rate. Our measurements in applied field examine how the static and dynamic magnetic fields contribute to the muon spin relaxation in the different temperature regimes. Finally, we discuss how our results on Bi$_2$Ir$_2$O$_7$ relate to those on other pyrochlore iridates and constrain the possible ground states for this material.

Polycrystalline samples of Bi$_2$Ir$_2$O$_7$ were synthesized from stoichiometric quantities of IrO$_2$ and Bi$_2$O$_3$ using a solid state reaction technique in an oxygen rich atmosphere, as described in Ref.~\onlinecite{qijpcm} where the characterization of the samples is detailed.
In our $\mu$SR experiments, carried out using the MuSR spectrometer at the ISIS Facility, UK, spin-polarized positive muons ($\tau_{\mu} = 2.2~\mu$s, $\gamma_{\mu} = 2\pi \times 135.5$~MHzT$^{-1}$) were implanted into the randomly-oriented polycrystalline sample, which was mounted in a $\sim 2$~cm$^2$ silver foil packet placed on a silver backing plate. 
The spin polarization of the implanted muons is measured by recording the asymmetry, $A(t)$, between the positron count rates in detectors placed on opposite sides of the sample, using the fact that positrons are most likely to be emitted in the direction of the muon's spin polarization when it decays.
Muons stop within the sample at interstitial sites or near electronegative atoms and if they experience magnetic fields perpendicular to their spin direction they will precess. If the magnetic fields are coherent between muon stopping sites then oscillations in the decay asymmetry may be observed. Magnetic fields parallel to the muon spin do not give rise to oscillations and depolarization only occurs if the fields are fluctuating. For randomly-oriented polycrystalline samples with bulk long-range magnetic order this gives rise to a $2:1$ ratio between oscillating and non-oscillating components of the decay asymmetry. 
The spins of muons stopping in the silver around the sample are depolarized much more slowly than those implanted in the sample and contribute a background to the asymmetry data that can be assumed constant. 
Measurements were carried out in zero applied magnetic field to look for spontaneous fields within the sample and in magnetic fields parallel and perpendicular to the initial muon spin direction to probe the distribution of magnetic fields within the sample.

\begin{figure}[t]
\includegraphics[width=0.9\columnwidth]{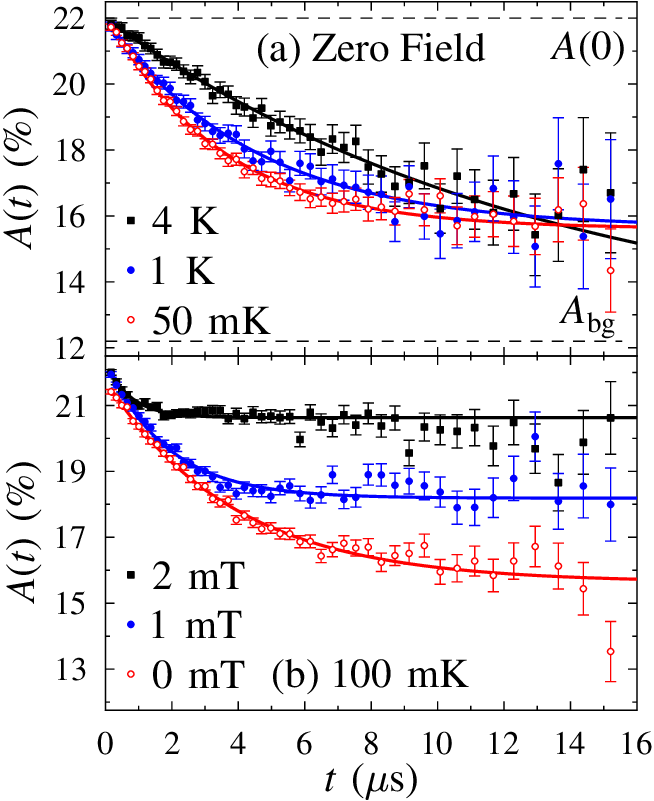}
\caption{ 
(Color online) 
$\mu$SR data with fits to eq.~\ref{zffit}.
(a) Zero field: Solid lines show fits to the data described in the text and dashed lines denote the initial, $A(0)$, and background, $A_{\rm bg}$, asymmetry values.
(b) $100$~mK: The apparent increase in the baseline is due to the decoupling shown in Fig.~\ref{musrlf}~(b).
\label{musrdata}}
\end{figure}

Raw muon decay asymmetry data measured in zero applied field are shown in Fig.~\ref{musrdata}(a). Data analysis was carried out using the WiMDA program.~\cite{wimda} At all measured temperatures from $50$~mK to $10$~K the muon relaxation was described effectively by an exponential form:
\begin{equation}
A(t) = A_{\rm bg} + A_{\rm s} [(1-f) \exp(-\lambda t) + f] .
\label{zffit}
\end{equation}
The values of the constant background $A_{\rm bg} = 12.2$~\% and sample $A_{\rm s} = 9.8$~\% terms were fixed using the data recorded above $2.5$~K with $f = 0$. No oscillations were evident in the zero field data. This function provides a simple way of describing our zero field data that is effective across the whole measured range and shows clearly where changes occur in the form of the data. However, as we describe below, it may well be a simplification of a more complicated underlying relaxation that is not directly apparent in the zero field measurements because of the unusually slow relaxation rates.

\begin{figure}[t]
\includegraphics[width=0.9\columnwidth]{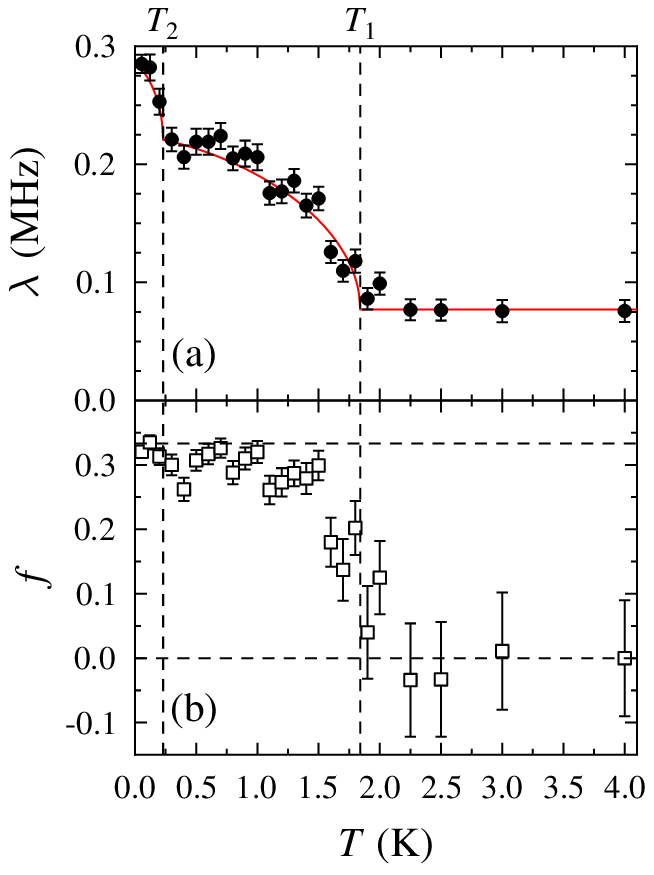}
\caption{ 
(Color online) 
Parameters derived from fitting equation~\ref{zffit} to the zero field $\mu$SR data shown in Fig.~\ref{musrdata}.
(a) Relaxation rate $\lambda$. The solid line is a guide to the eye.
(b) Fractional amplitude of the non-relaxing asymmetry, $f$. Dashed lines show transition temperatures and limiting values of $f$.
\label{musrzf}}
\end{figure}

The parameters $\lambda$ and $f$ obtained from fitting equation~\ref{zffit} to the zero field data are shown in Fig.~\ref{musrzf}. Above $2$~K both parameters take constant values. We find two temperatures at which $\lambda$ increases steeply with decreasing temperature. The guide to the eye shown in Fig.~\ref{musrzf}(a) is the sum of two terms of the form $\lambda_i (1-(T/T_i)^{1.5})^{0.5}$, each of which is taken to be zero above the respective $T_i$, and a residual relaxation, $\lambda_{\rm r}$, that is likely to be due to paramagnetic electronic moments and the distribution of fields from the nuclear moments. This allows us to estimate by least-squares fitting the transition temperatures $T_1 = 1.84(3)$ and $T_2 = 0.23(4)$~K, which are not particularly sensitive to the choice of the exponents in the power law. 
The corresponding relaxation rates are $\lambda_1 = 0.15(1)$, $\lambda_2 = 0.06(1)$, and $\lambda_{\rm r} = 0.081(4)$~MHz.

The non-relaxing fraction, $f$, of the signal from the sample changes over the region $1.5$ to $2.25$~K and is shown in Fig.~\ref{musrzf}(b). Below $1.5$~K, $f=1/3$ within experimental error. This suggests that spontaneous, static magnetic fields develop throughout the volume of the sample below the transition at $1.84$~K. It is unusual to observe a simple exponential relaxation in a static magnetic state, where one would expect either oscillations due to well-defined fields at muon stopping sites or a Kubo-Toyabe-like function due to a broad distribution of magnetic fields at the stopping sites. 

\begin{figure}[t]
\includegraphics[width=0.9\columnwidth]{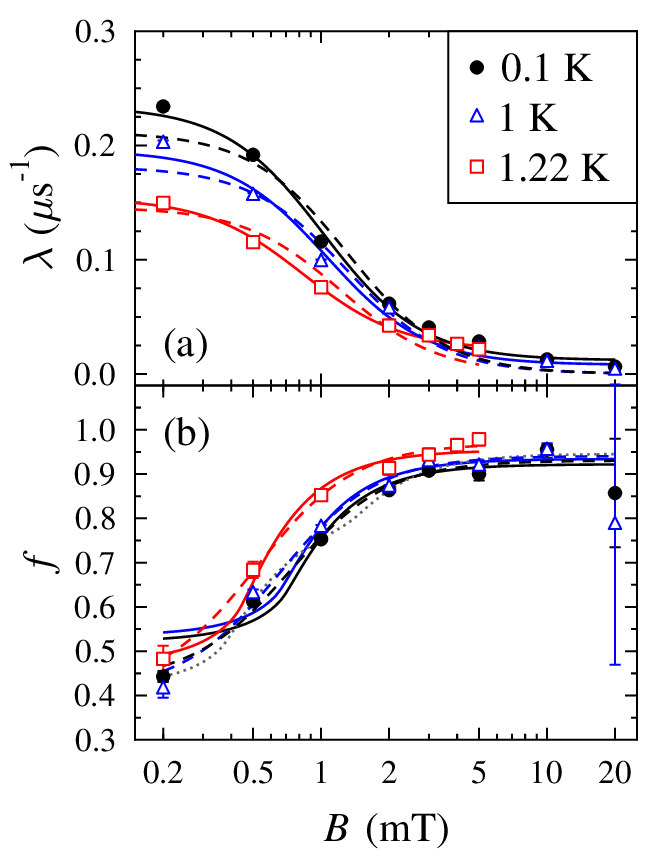}
\caption{ 
(Color online) 
Parameters derived from two approaches to fitting the muon data in a magnetic field $B$ applied parallel to the spins of the implanted muons.
(a) Relaxation rate $\lambda$ with fixed $f$. The solid and dashed lines represent fits to eq.~\ref{redfield} with and without a field-independent offset value $\lambda_0$.
(b) Non-relaxing fraction, $f$, with $\lambda$ allowed to vary. The solid and dashed lines represent fits to equations~\ref{prattfit} and~\ref{muonium} respectively. The dotted line is discussed in the text.
Each fitting methodology is described in the text.
\label{musrlf}}
\end{figure}
To attempt to elucidate the muon relaxation mechanism and underlying magnetic state in Bi$_2$Ir$_2$O$_7$ we carried out measurements with a magnetic field applied parallel to the initial muon spin direction, the longitudinal field (LF) geometry. These were performed at $0.1$, $1$, and $1.22$~K. Increasing the applied field gradually decouples the muon relaxation from the field distribution it experiences from its surroundings, providing information on the form of that field distribution. If the relaxing amplitude of the signal [$(1-f)$ in eq.~\ref{zffit}] is field independent this suggests that dynamic local magnetic fields depolarize the muon spin, whereas a static field distribution results in a growth of the constant part of the signal ($f$ in eq.~\ref{zffit}) as the applied field exceeds the characteristic static field. The relaxation rate $\lambda$ can also show informative trends with the applied field $B$.

We used two approaches to analyse the LF-$\mu$SR data shown in Fig.~\ref{musrdata}(b), firstly considering a situation where magnetic dynamics contribute significantly to the observed relaxation, and secondly concentrating on a quasistatic scenario. In the first of these we fixed the amplitudes of the relaxing and constant components of eq.~\ref{zffit} to their zero field values, leaving $\lambda$ as the only free parameter. The field dependence of $\lambda$ at the three measured temperatures is shown in Fig.~\ref{musrlf}(a). The relaxation rate decreases with applied field as is conventionally expected and we fitted the field dependence using a modified version of Redfield's equation, as used in Ref.~\onlinecite{disseler12yyb}:
\begin{equation}
\lambda = \frac{2\gamma^{2}_{\mu}\Delta^{2}\tau}{1+\gamma^{2}_{\mu}B^{2}\tau^{2}} + \lambda_0.
\label{redfield}
\end{equation}
The constant $\gamma_{\mu}$ is the muon gyromagnetic ratio, $\Delta$ describes the width of the field distribution, and $\tau$ is the characteristic timescale for the spin fluctuations experienced by the muons. We found that the quality of our fit to the field dependence was improved considerably by adding a field independent offset $\lambda_0 \sim 0.01$~MHz, as was found for Yb$_2$Ir$_2$O$_7$ [$\lambda_0 = 0.05(2)$~MHz],~\cite{disseler12yyb} and as can be seen in Fig.~\ref{musrlf}(a). However, $\lambda_0$ does not make a large difference to the extracted values of $\Delta/\gamma_{\mu} = 0.35(2)$~mT and $\tau = 1.1(1)~\mu$s for $0.1$ and $1$~K, with $\Delta$ decreasing and $\tau$ increasing slightly at $1.22$~K. $\Delta$ is thus an order of magnitude smaller in Bi$_2$Ir$_2$O$_7$ than in Yb$_2$Ir$_2$O$_7$ and $\tau$ is three orders of magnitude larger.~\cite{disseler12yyb} To get an exponential muon spin relaxation due to motional narrowing requires $(\Delta\tau)^{-1} \gtrsim 3$. The value obtained from the fitted parameters is $3.05$, suggesting internal consistency. However, it is not possible to determine these parameters independently from the zero field data and the change of $f$ around 1.84~K is not obviously compatible with this fitting function since $f$ should be zero at all temperatures. This analysis therefore excludes strongly dynamic local fields and suggests that the field dependence of $\lambda$ must result from a quasistatic distribution of local fields, as we discuss in more detail below.

The second approach was to allow the amplitude of the constant component of eq.~\ref{zffit}, $f(B)$, to vary (with $\lambda$ as a free parameter) and the values of $f(B)$ are shown in Fig.~\ref{musrlf}(b). We took two different models for the field dependence of $f$ to compare with the data: a single characteristic field $B_c$,~\cite{prattjpcm}
\begin{equation}
f(B) = f_0 + F \Big( \frac{3}{4} - \frac{1}{4b^2} + \frac{(b^2 - 1)^2}{8b^3}\log{\Big{\vert}\frac{b+1}{b-1}\Big{\vert}} \Big) ,
\label{prattfit}
\end{equation}
where $b = B/B_c$, and a quadratic-type decoupling function appropriate for a Lorentzian distribution of fields,~\cite{prattjpcm}
\begin{equation}
f(B) = f_0 + F \frac{b^2}{1+b^2}.
\label{muonium}
\end{equation}
We note that the quadratic-type decoupling appears to describe the data better than decoupling from a single characteristic field and this suggests that there is a distribution of fields around the characteristic field. To further model this decoupling we fitted a weighted sum of two terms of the form of eq.~\ref{prattfit} with different characteristic fields. This is shown as the dotted line in Fig.~\ref{musrlf}(b) for $0.1$~K, with characteristic fields of $0.37(3)$ and $1.4(2)$~mT, and a fractional amplitude of $0.31(4)$ associated with the higher field.
The values of $B_c$ coming from these three fits are again very similar, with $B_c =0.68(9)$~mT from eq.~\ref{prattfit}, $0.73(3)$~mT from eq.~\ref{muonium}, and a weighted average of $0.69(9)$~mT from the two field model. Fitting the data to a Lorentzian Kubo Toyabe function with this width proves successful and this can be used to fit the data up to around 1.84~K, where the amplitude of this term and the background change similarly to those of eq.~\ref{zffit}, giving further grounds for identifying a transition at $T_1$. 

Transverse field measurements ($B=2$~mT) were performed to search for more strongly magnetic regions within the sample, but no discernible temperature dependence was evident in the asymmetry of the oscillations and the relaxation rate was compatible with the zero field measurements. Thus we have no evidence for strongly magnetic regions dispersed within the sample.

Given the small magnetic field experienced by the implanted muons, $B_c = 0.7$~mT, Bi$_2$Ir$_2$O$_7$ seems to be quite exceptional compared to the Pr ($9$~mT),~\cite{maclaughlin10} Nd ($67$~mT),~\cite{disseler12nd} Y ($95$~mT),~\cite{disseler12yyb} Eu ($99$~mT),~\cite{zhao11} and Yb ($113$~mT)~\cite{disseler12yyb} pyrochlore iridates. The iridium moment size in Eu$_2$Ir$_2$O$_7$ was estimated from muon measurements~\cite{zhao11} to be $\mu_{\rm Ir} \approx 1.1~\mu_{\rm B}$ whereas in Nd$_2$Ir$_2$O$_7$ the absence of magnetic Bragg peaks placed an upper bound on the moment size of $\leq 0.5~\mu_{\rm B}$.~\cite{disseler12nd} Taking the magnetic field experienced by the muon to scale with the moment size suggests for Bi$_2$Ir$_2$O$_7$ that $0.005 < \mu_{\rm Ir}/\mu_{B} < 0.01$, the upper limit being quite close to the saturation magnetization at $1.7$~K of $0.013~\mu_{\rm B}$/Ir.~\cite{qijpcm} Because of the exceptionally small moment it is hard to elucidate whether persistent fluctuations of an ordered state are present, as was suggested from $\mu$SR data for Eu$_2$Ir$_2$O$_7$,~\cite{zhao11} BaIrO$_3$,~\cite{brooks05} and Sr$_2$IrO$_4$.~\cite{franke} The low transition temperatures of Bi$_2$Ir$_2$O$_7$ are also exceptional in this series, with most members showing magnetic transitions at temperatures above 100~K, except Nd$_2$Ir$_2$O$_7$ ($T_{\rm N} = 8$~K)~\cite{disseler12nd} and Pr$_2$Ir$_2$O$_7$ where long-range magnetic ordering has not been found down to $30$~mK.~\cite{maclaughlin10} 

Since the lattice constant of Bi$_2$Ir$_2$O$_7$ is between those of Eu$_2$Ir$_2$O$_7$ and Nd$_2$Ir$_2$O$_7$ we might anticipate its properties would be similarly intermediate, but this is not the case.
Alternatively, we can draw a comparison with metallic Pr$_2$Ir$_2$O$_7$, which has the next largest $A$-site ionic radius in the series, shows more features in common, but no Ir-moment magnetic ordering has been found in Pr$_2$Ir$_2$O$_7$ in spite of detailed studies showing other novel features in its magnetic properties.~\cite{maclaughlin10,nakatsuji06,machida07,machida10} 
The reduced magnetic moment of the iridium ions and reduced ordering temperature for Bi$_2$Ir$_2$O$_7$ may therefore be due to the large ionic radius of Bi, but may also relate to the lack of hyperfine enhancement from rare earth moments~\cite{chen12xxx} or a greater degree of itinerancy or closer proximity to a quantum critical point.~\cite{qijpcm}
A recent theoretical study of pyrochlore iridates~\cite{witczakkrempa13} identifies a metallic antiferromagnetic phase with significantly reduced magnetic moments and ordering temperatures compared to the insulating antiferromagnetic phase, but a similar all-in/all-out magnetic structure. Such a description appears wholly compatible with our results and the small, negative $\theta_{\rm CW} = -2.3$~K found by Qi {\em et al.}~\cite{qijpcm}
We also note that only a small proportion of the magnetic moment determined from the Curie-Weiss fits, $0.1~\mu_{\rm B}$,~\cite{qijpcm} appears to be associated with the quasistatic magnetism, determined either from the saturation magnetization measured in the bulk~\cite{qijpcm} or the $\mu$SR results presented here. 

Finding two transitions in Bi$_2$Ir$_2$O$_7$ at low temperature draws another parallel with the two transitions seen in other pyrochlore iridates such as Nd$_2$Ir$_2$O$_7$~\cite{disseler12nd} and Y$_2$Ir$_2$O$_7$.~\cite{disseler12yyb} These two transitions were suggested to result from frustration preventing magnetic ordering above the temperature at which it is relieved by a metal-insulator transition, as opposed to the sharp onset of long-range order in insulating pyrochlore iridates.~\cite{zhao11,disseler12yyb} Resistivity measurements below $1.7$~K have not been carried out so we cannot complete this comparison, although the similarity of the field-dependent results at $0.1$ and $1$~K suggests that the magnetic states on either side of the transition at $0.23$~K are similar and therefore a dramatic change in the electronic properties at that temperature is unlikely.

In conclusion, we have observed a magnetic transition in Bi$_2$Ir$_2$O$_7$ at 1.84(3)~K below which the muon relaxation rate sharply increases and find evidence for a further transition at 0.23(4)~K.
The associated change in the $\mu$SR baseline asymmetry suggests that the upper transition represents the material entering a quasistatic magnetic state throughout its entire volume. The lower transition can be associated with a marked upturn in the low-temperature specific heat below this temperature,~\cite{qijpcm} but its physical origin remains unclear. From the magnetic field dependence of the $\mu$SR data we estimate the fields experienced by muons in Bi$_2$Ir$_2$O$_7$ to be two orders of magnitude smaller than in comparable iridate pyrochlores and this suggests a moment size, $\sim 0.01~\mu_{\rm B}$/Ir, compatible with the saturation magnetization previously determined.~\cite{qijpcm} Comparison with theoretical work~\cite{witczakkrempa13} suggests that the low temperature state is likely to be a weakly antiferromagnetic metal, which has not previously been observed in the pyrochlore iridate series.

We acknowledge the provision of beamtime at the ISIS Facility by STFC, thank EPSRC (UK) for financial support, and thank William Witczak-Krempa for helpful discussions. The work at the University of Kentucky was supported by NSF through grants DMR-0856234 and EPS-0814194.  


\end{document}